# Crystal structures of Fe-gluconate

Łukasz Gondek[*] and Stanisław M. Dubiel

*AGH University of Science and Technology, Faculty of Physics and Applied Computer Science, Mickiewicza 30, 30-059 Kraków, Poland*

**Abstract**

Fe-gluconate, $Fe(C_6H_{11}O_7)_2 \cdot xH_2O$ is a well-known material widely used for iron supplementation, in food industry as a coloring agent, in cosmetic industry for skin and nail conditioning, and metallurgy. Despite of wide range of applications its physical properties were not studied extensively. In this study, Fe-gluconate with three different amount of water viz. $x$=2 (fully hydrated), $0 < x < 2$ (intermediate) and $x$=0 (dry) was investigated by means of X-ray diffraction (XRD) and Mössbauer spectroscopic (MS) methods. The former in the temperature range of 20-300 K, and the latter at 295 K. Based on the XRD measurements crystallographic structures were determined: monoclinic (space group *I2*) for the hydrated sample and triclinic (space group *P1*) for the dry sample. The partially hydrated sample was two-phased. Unit cells parameters for both structures show strong, very complex and non-monotonic temperature dependences. Mössbauer spectroscopic measurements gave evidence that iron in all samples exist in form of Fe(II) and Fe(III) ions. The amount of the latter equals to ~30% in the hydrated sample and to ~20% in the dry one.

**Keywords:** Fe-Gluconate; Crystal structure; Mössbauer spectroscopy; X-ray diffraction

*corresponding author: lgondek@agh.edu.pl

# 1. Introduction

Fe-gluconate is a metallorganic compound (a salt of the gluconic acid), whose chemical formula can be written as $Fe(C_6H_{11}O_7)_2 \cdot xH_2O$. The majority of iron is in its Fe(II) (or $Fe^{2+}$) form, yet a minor fraction ($\leq \sim 15\%$) of iron exists as Fe(III) ($Fe^{3+}$) ions as revealed using Mössbauer spectroscopic studies [1,2,3]. Fe-gluconate has been commonly used in the pharmacological industry for a production of supplements, e. g. Apo® ferrous, Apotex®, Ascofer®, Fergon®, Ferralet®, Simron®, to name just a few, applied in the treatment of iron deficiency anemias. The Fe-gluconate is also used in food industry as a coloring agent (registered as E579) and in cosmetic industry for skin and nail conditioning (CAS number: 299-29-6). Noteworthy, it was also used in the metallurgical industry as an effective inhibitor for a carbon steel [4], and gluconate-based electrolytes were also successfully applied to electroplate various metals [5] or alloys [6]. It is well known that an iron intake by humans depends on a number of factors, one of them being its redox state (ferrous or ferric) e. g. [7]. The bioavailability of Fe(II) ranges between 10 and 15% and is 3-4 times higher than the one of Fe(III) [8].

The purity of the Fe-gluconate is thus of great importance as far as its medical and cosmetic applications are concerned. However, it seems that not chemical only but also structural purity should be considered. Therefore, the question whether the ferrous and the ferric components origin from the same crystallographic structure was raised. Astonishingly, only the structure of hydrous Fe-gluconate has been just recently determined based on the synchrotron powder diffraction data [9]. No evidence of structural changes varying with the water content was reported till date.

The aim of the present study was to investigate structural and electronic properties of the hydrated and dry samples of the Fe-gluconate.

# 2. Experimental

Three samples of Fe-gluconate, $Fe(C_6H_{11}O_7)_2 \cdot xH_2O$, with different amount of water viz. hydrated (H) $x=2$, dry (D) $x=0$, and intermediate, called hereafter untreated (U), $0 < x < 2$ were investigated by means of X-ray diffraction (XRD) and Mössbauer spectroscopic (MS) techniques. The XRD measurements were performed using Panalytical Empyrean powder diffractometer with Cu K$_\alpha$ radiation. Oxford Cryosystem PheniX closed-cycle cryostat was used for low temperature XRD (20-300K). During the low-temperature measurements the sample position was maintained using a motorized stage with calibration curve. For high-temperature studies (30 - 160°C) Anton Paar HTK 1200N chamber was used. The sample position was also calibrated against the temperature displacements. The diffraction patterns were indexed and refined using FullProf Suite software [10].

Molecular mechanics simulations were made using SAMSON package [11] with usage of Universal Force Field [11, 12].

The Mössbauer spectra were recorded at room temperature (295 K) in a transmission geometry using a standard (Wissel GmbH) spectrometer and a drive working in a sinusoidal mode. Each spectrum was recorded in 1024 channels and 14.4 keV gamma rays were supplied by a $^{57}$Co/Rh source whose activity enabled recording a statistically good spectrum within a 2-3 days run.

## 3. Results and discussion
*3.1 Ambient temperature XRD studies*

X-ray diffraction measurements revealed that the Fe-gluconate exhibited two crystal structures depending on the water content. The diffraction pattern for a dry sample can be indexed by the triclinic unit cell with parameters listed in Fig. 1. On the other hand, after hydrating the sample the corresponding XRD pattern could be fully indexed and refined using *I2* space group with lattice parameters close to those reported in the literature [9].

Few conformations (Conf. 1-3) of Fe-gluconate, as derived from molecular mechanics optimization, are presented in the Fig. 1, as well. Analyses of those conformations can suggest orientation of the molecules in the crystal unit cell. It seems apparent that the structure of a hydrated Fe-gluconate is related to Conf. 1, which longitudinal dimension 19.6 Å correlates quite well with the lattice parameter *a* = 19.86 Å of the crystal unit cell. However, the drying process turns the structure into a lower triclinic symmetry. In this case, Conf. 2 appears to be the most suitable as its longitudinal dimension (17.2 Å) is in agreement with the lattice parameter *a* = 17.57 Å. The linking O-Fe-O bridge appears to be quite stiff and behave very similar for all conformations.

To study the transition between hydrated and dry phase of Fe-gluconate the high-vacuum (6·10$^{-7}$ mbar) was applied to Anton-Paar chamber, which temperature was set at 350 K. During exposing to vacuum, the XRD patters were collected in intervals of 15 min. As apparent from Fig. 2, the hydrated sample with the monoclinic structure was turning into the triclinic structure. During drying, coexistence of both structures (monoclinic and triclinic) was evidenced. One may notice that the triclinic structure is changing significantly during the process, as can be inferred from a splitting of the reflection at 8.5° of the 2θ-range. Finally, after 45 min. the sample was entirely dry exhibiting only the monoclinic phase. It should be underlined that the hydrated Fe-gluconate will deteriorate into two phases (hydrated an dry) after couple of weeks at ambient conditions.

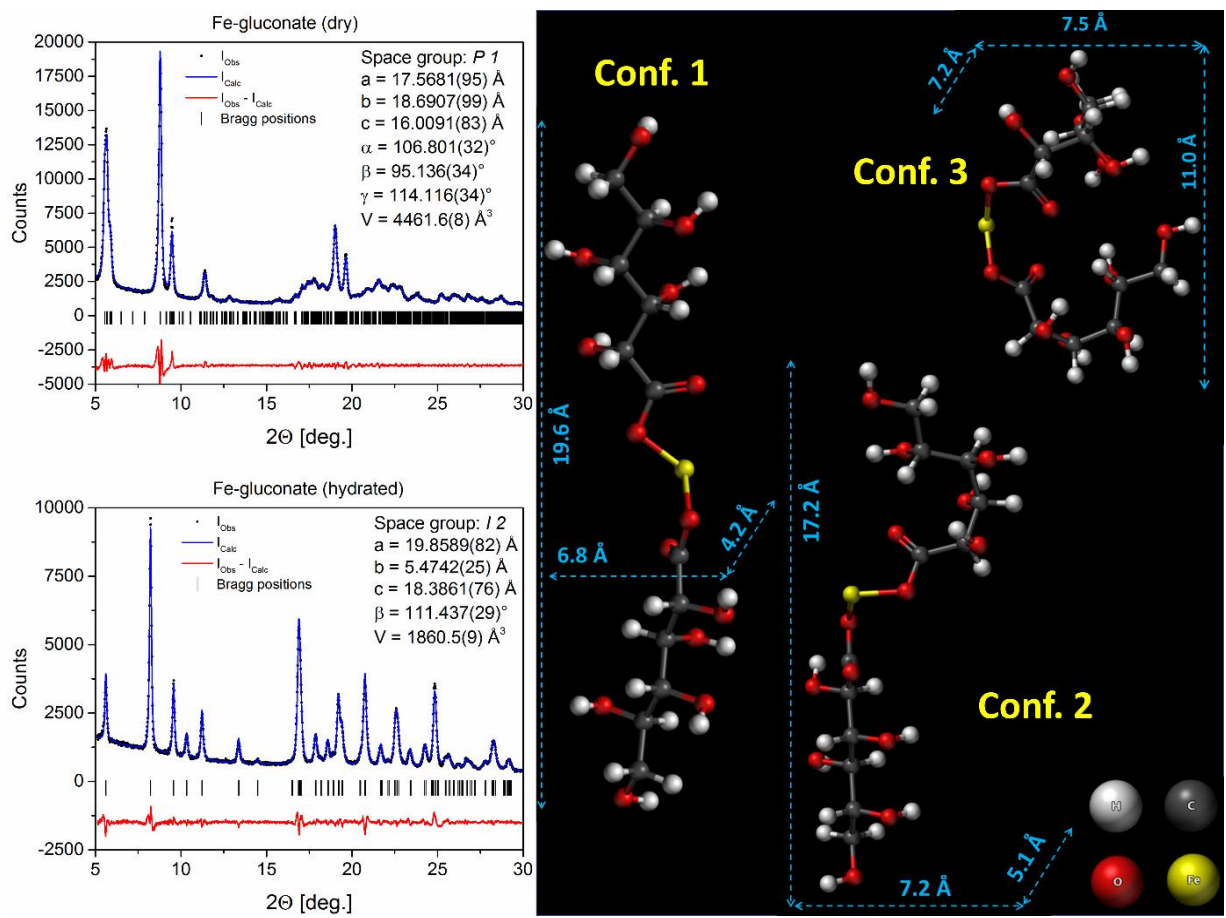

Fig. 1 Ambient conditions XRD patterns of dry (upper part) and hydrated (bottom part) samples of the Fe-gluconate. The diffractograms were refined using the La Bail method; results of the refinements are given on each diagram. On the right-hand side few of possible conformations of the Fe-gluconate are presented with estimated dimensions.

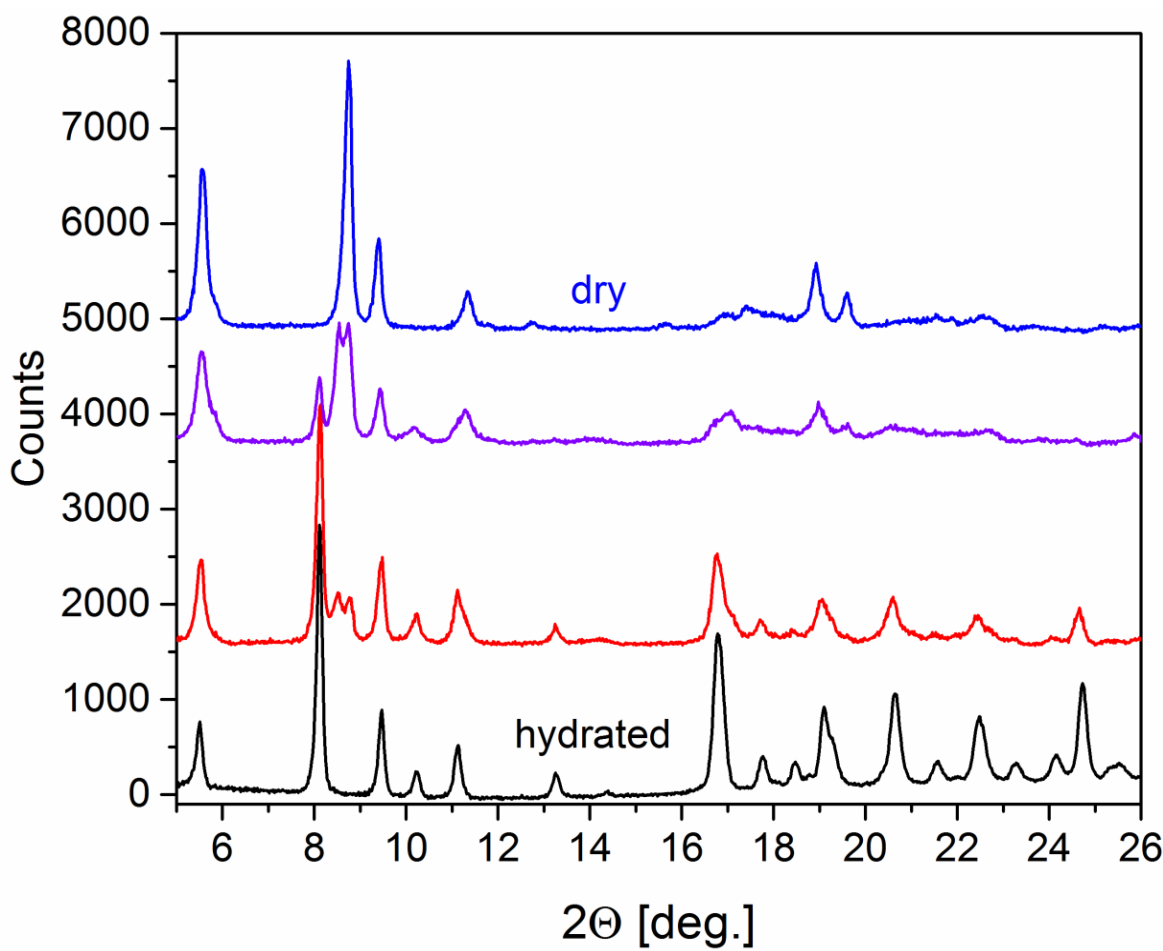

Fig. 2 Evolution of XRD patterns for the Fe-gluconate during drying under dynamic high-vacuum conditions at 350K. Time of the drying was 45 min.

*3.2 Non-ambient temperature XRD studies*

Low temperature XRD patterns for the hydrated Fe-gluconate are presented in Fig. 3. The room-temperature monoclinic structure is retained down to 20 K. However, below 80 K some reflections are changing its position and intensities significantly, as presented in the Fig. 3 for the -202 reflection. However, the crystallographic structure still can be described by the same *I*2 space group.

La Bail refinement shows that lattice parameters behave anomalously at low temperatures, as presented in Fig. 4. Namely, one could expect a kind of saturation at low temperatures, according to the Debye formula for thermal expansion. Contrary to this expectation, the *a* and *c* lattice parameters are evidenced to decrease significantly below 140 K without an expected plateau at lowest temperatures. The behavior of the *b* lattice parameter is much more complex. It exhibits anomalous behaviors at around 60-80 K, 200 K and 310 K. The first anomaly can be apparently associated with aforementioned changes of the -202 reflection intensity below 80 K.

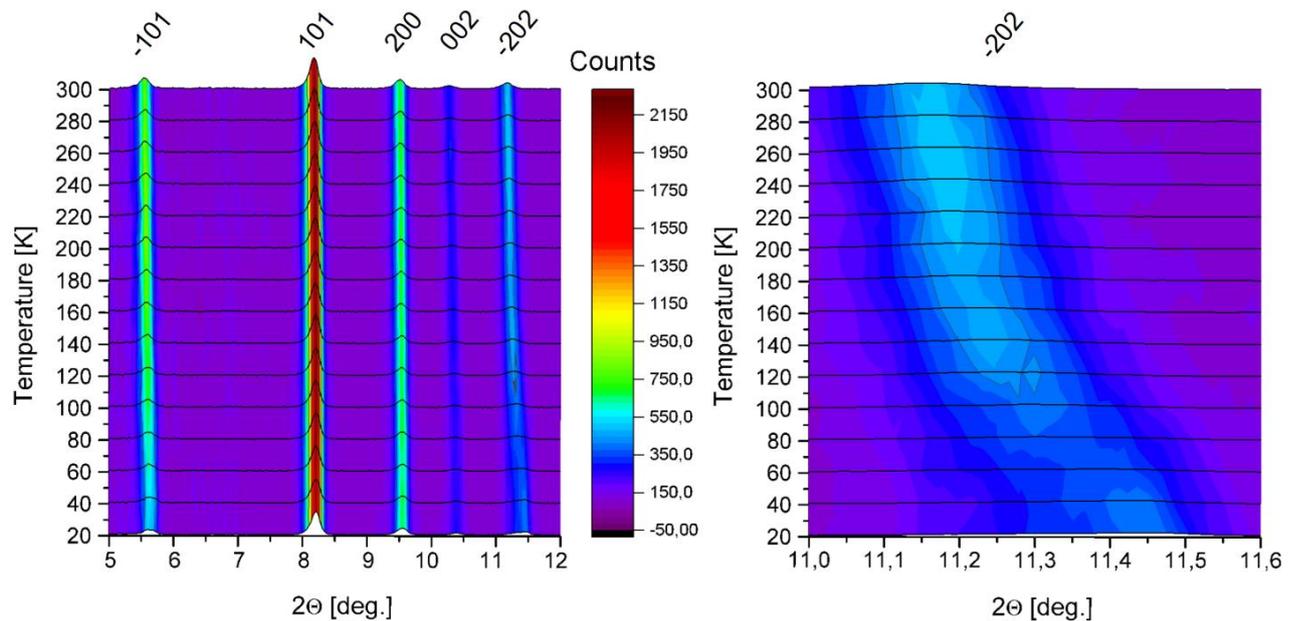

Fig. 3 Low temperature XRD patterns for the hydrated Fe-gluconate. The -202 reflection is magnified to highlight some structural changes.

All lattice parameters display a significant decrease above 374 K, which is connected to releasing water from the crystal structure of the hydrated Fe-gluconate. The phase transition from the monoclinic (hydrated) to the triclinic (dry) structure of the Fe-gluconate is depicted in Fig. 5. Around that temperature, the reflections of the monoclinic phase are shifted towards higher 2θ-angles, marking a collapse of the structure. In between, i.e. 374-389 K, some complex intermediate phase is formed, while at 394 K the triclinic (dry) phase is finally stabilized. The triclinic phase is stable up to 424 K, when reflections are rapidly disappearing yielding loose of the long-range ordered structure (see the Fig. 5).

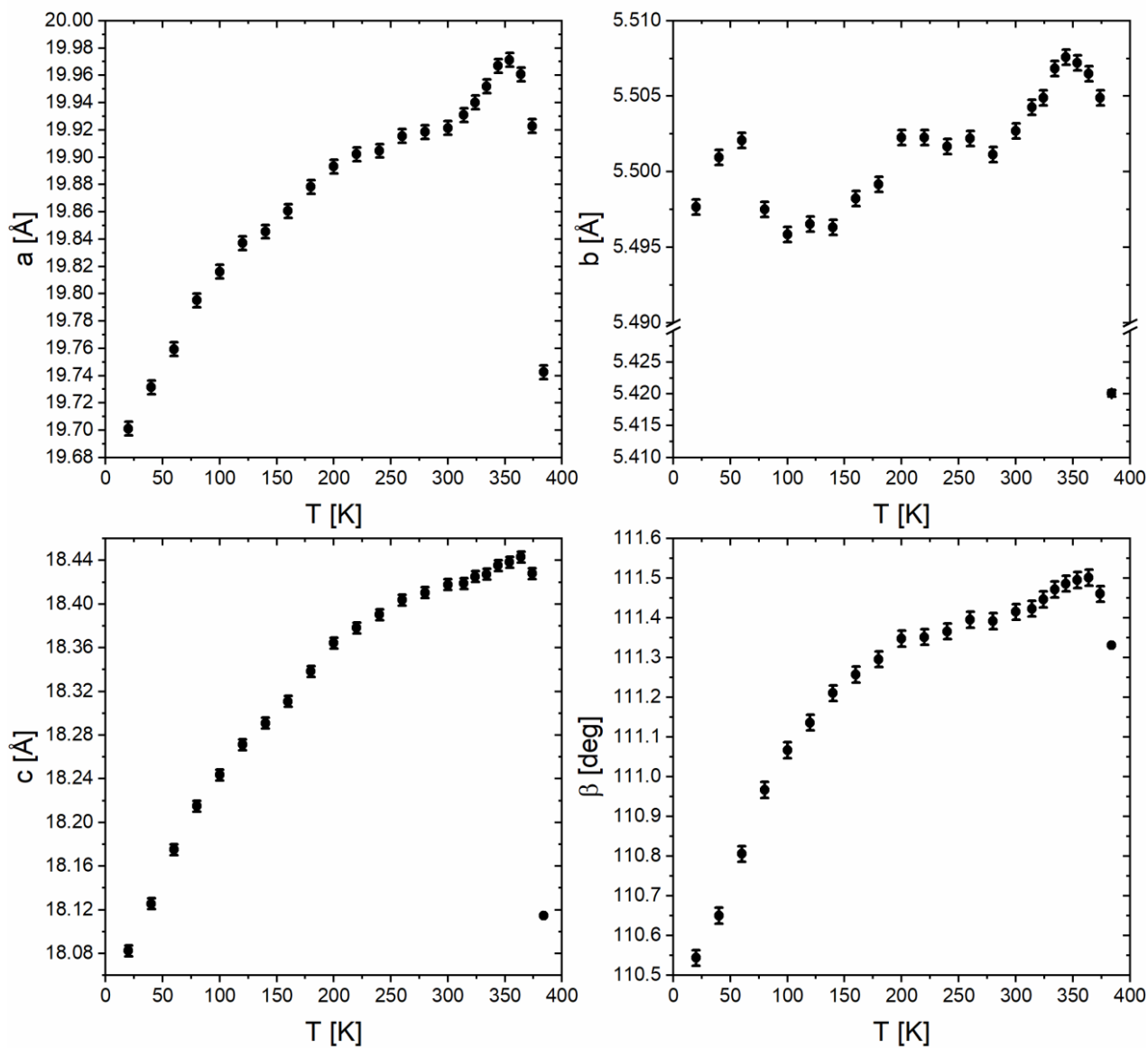

Fig. 4 Results of the La Bail refinement of the monoclinic (hydrated) Fe-gluconate phase in full range of stability.

One may notice a significant difference between drying the sample at room temperature under the high vacuum conditions, where a coexistence of the hydrated and dry phases was evidenced (see the Fig. 2), and exposing the sample to heating, when no coexistence of the two phases is present.

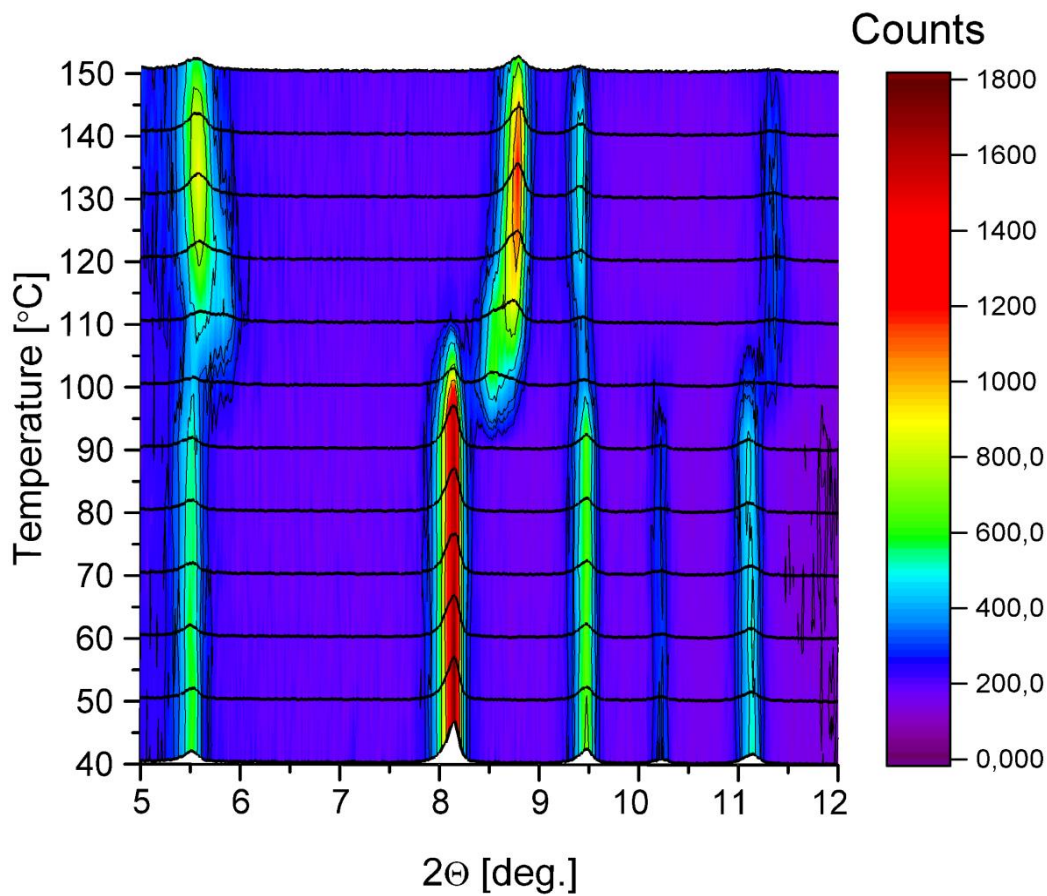

Fig. 5 High temperature XRD patterns of the hydrated Fe-gluconate (heating in air).

Also the dry Fe-gluconate was measured at low temperatures. Results of the refinement in this case are given in Fig. 6. In the entire temperature range (20-300K) the monoclinic phase was stable. However, changes of the lattice parameters presented in the Fig. 6 show a very complex behavior. Namely, the lattice parameter *a* decreases quite smoothly with a pronounced plateau at low temperatures, while the *b* and the *c* parameters shows anomalous decrease of the value with rising temperature. This behavior is somewhat similar to Metal-Organic Frameworks (MOFs), where vibration of rigid metal-complexes leads to a shrinking of the lattice with temperature increase. Apart from the above behavior, also $\alpha$, $\beta$ and $\gamma$ angles exhibit non-trivial temperature dependences as displayed in Fig. 6. Comparison between the hydrated and the dry specimens of the Fe-gluconate leads to conclusion that a lattice dynamics of this compound is extremely complex.

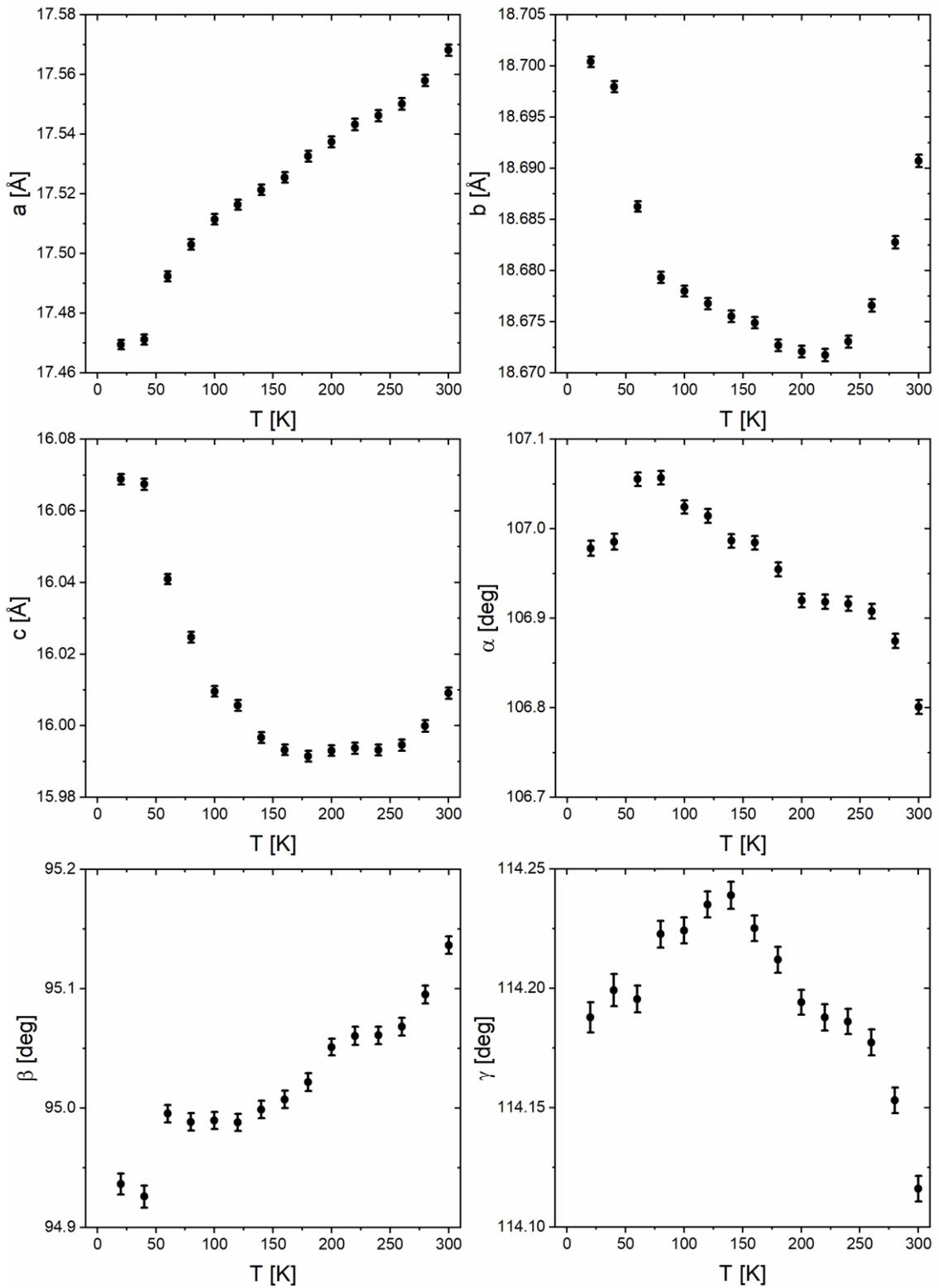

Fig. 6 Results of the La Bail refinement of the triclinic (dry) Fe-gluconate phase between 20-300K.

*3.3 Ambient temperature Mössbauer-effect studies*

The collected Mössbauer spectra, examples of which are shown in Fig. 7, were analyzed using a least-squares fitting procedure.

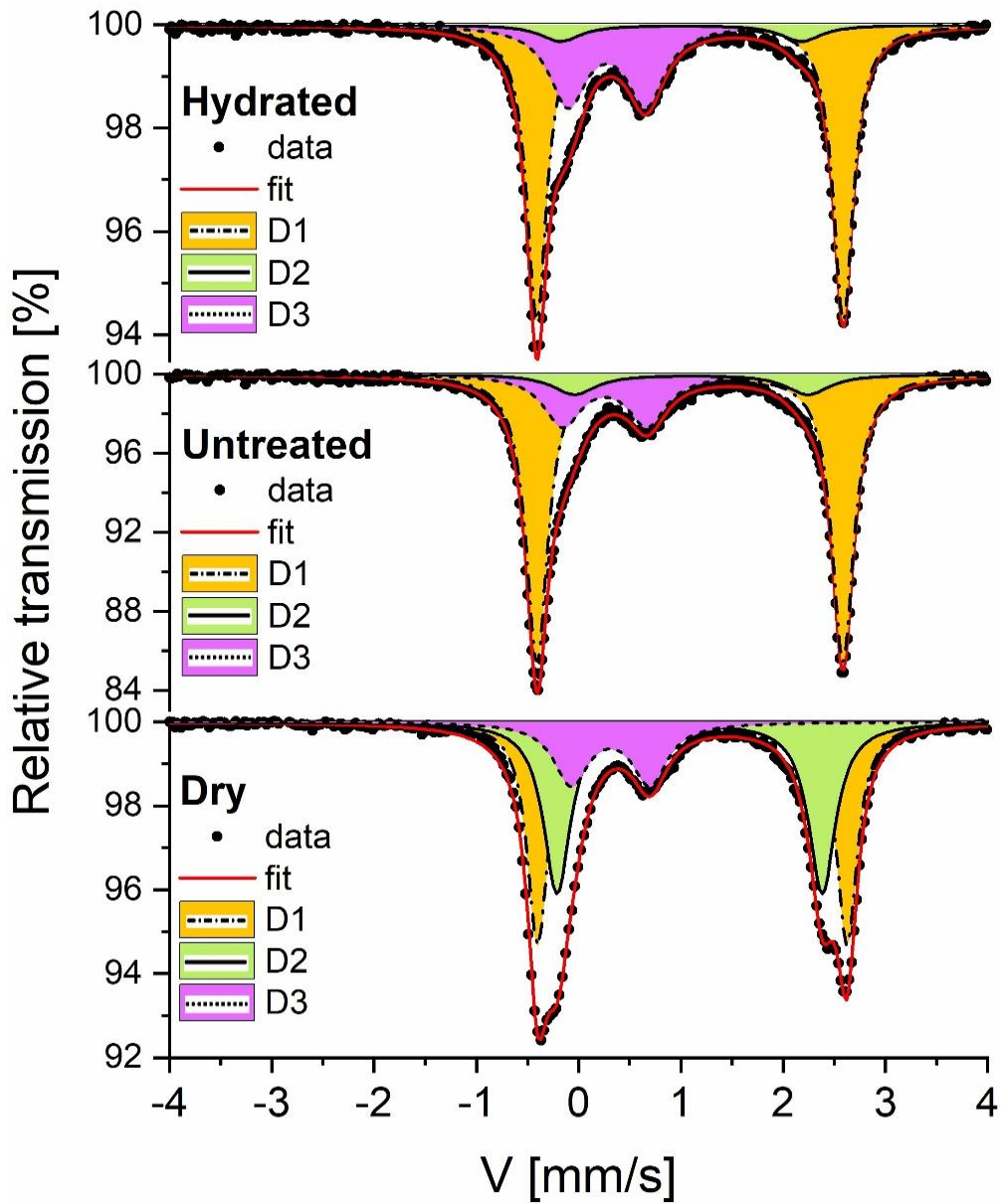

Fig. 7 $^{57}$Fe Mössbauer spectra recorded at room temperature for the investigated samples. The filling colors representing D1, D2 and D3 contributions match colors of the results presented in Fig. 8.

The spectra could be decomposed using three doublets, D1, D2 and D3. The first two were associated with the outermost lines constituting a major component of each spectrum. The D1 and D2 doublets can be associated to $Fe^{2+}$ contribution, while the D3 doublet was ascribed to a minor $Fe^{3+}$ component. The latter doublet with smaller quadrupole splitting and isomeric shift (characteristic of a high-spin $Fe^{3+}$ state) is visible between the two major lines. Isomer shifts (IS1, IS2, IS3), quadrupole splitting (QS1, QS2, QS3), line widths (G1, G2, G3) and relative contributions (A1, A2, A3) were treated as free parameters. Very good fits were obtained in terms of $\chi^2$ and misfit tests. Values of the best-fit spectral parameters of the three sub spectra (doublets) are presented in Fig. 8 and Table 1.

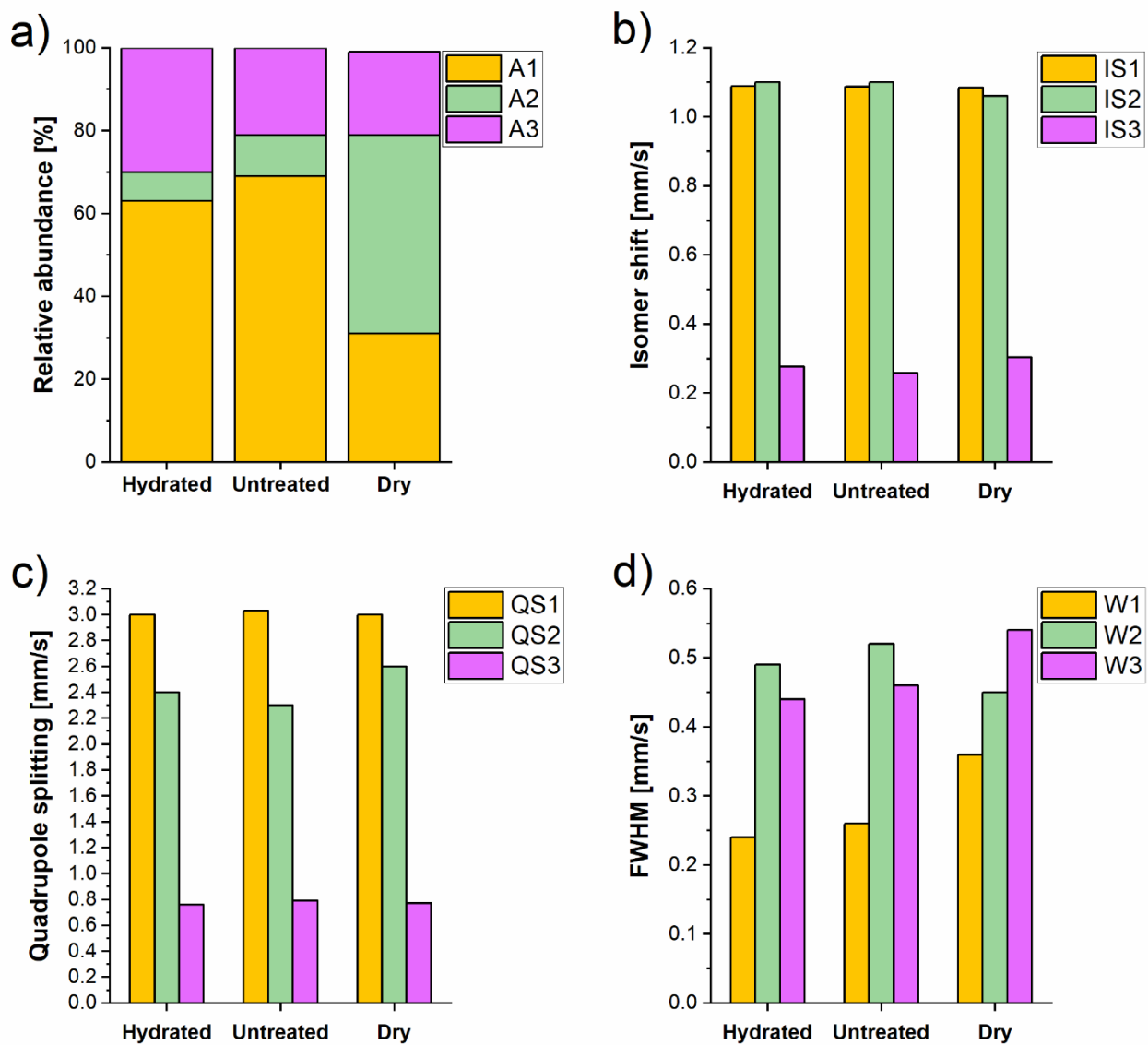

Fig. 8 Results of Mössbauer spectra analysis: a) Relative abundances (A); b) Isomer shifts (IS); c) Quadrupole splittings (QS); d) Full linewidths at half maximum (FWHM) of the three doublets evidenced the studied samples. Numbers 1 and 2 are related to the major Fe(II) phases (D1 and D2), while number 3 refers to the minor Fe(III) phase (D3).

Table 1 Values of the best-fit spectral parameters for evaluation of Mössbauer spectroscopy data using three doublets D1, D2 and D3. A1, A2, A3 – relative abundances; IS1, IS2, IS3 – isomer shifts; QS1, QS2, QS3 – quadrupole splittings, W1, W2, W3 - full linewidths at half maximum.

| Parameter | Hydrated | Untreated | Dry |
|---|---|---|---|
| A1 [%] | 63(1) | 69(1) | 32(1) |
| A2 [%] | 7(2) | 10(2) | 48(1) |
| A3 [%] | 30(1) | 21(1) | 20(1) |
| IS1 [mm/s] | 1.089(1) | 1.088(1) | 1.085(1) |
| IS2 [mm/s] | 1.10(1) | 1.10(1) | 1.06(1) |
| IS3 [mm/s] | 0.277(3) | 0.258(3) | 0.303(3) |
| QS1 [mm/s] | 3.00(5) | 3.03(5) | 3.00(5) |
| QS2 [mm/s] | 2.4(1) | 2.3(1) | 2.6(1) |
| QS3 [mm/s] | 0.76(8) | 0.79(8) | 0.77(8) |
| W1 [mm/s] | 0.24(1) | 0.26(1) | 0.36(1) |
| W2 [mm/s] | 0.49(5) | 0.52(5) | 0.45(5) |
| W3 [mm/s] | 0.44(2) | 0.46(2) | 0.54(2) |

*3.3.1 Relative abundances*

There is a striking difference between the abundances of 3 doublets in hydrated and the dry sample, as evidenced in Fig. 8a. Firstly, the relative amount of the ferric component is by 10% higher in the hydrated sample in comparison to the others. Secondly, the relative amount of the D2 component in the dry sample is by a factor of 7 higher than in the hydrated specimen. Surprisingly, the amount of the ferric phase in the untreated specimen is similar to the dry sample, while the amount of the D2 contribution is almost the same as in the hydrated specimen. This suggests that there is no simple superposition of hydrated and dried phases. Taking into account the XRD results presented in the Fig. 2 reviling that the dried phase changes during water extraction, it becomes understandable that changes occurring in drying triclinic phase are responsible for D2 abundance growth.

*3.3.2 Isomer shifts and quadrupole splitting*

The Fig. 8b gives evidence that the values of the isomer shift (charge-density) hardly depend on the amount of water in the Fe-gluconate. As clearly evidenced in the Fig. 8c also the values of the quadrupole splitting (electric field gradient) are fairly independent on the presence of water in the Fe-gluconate. Thus one can conclude that the electronic structure of the Fe-gluconate as seen by Fe ions (both ferrous and ferric) is not affected by water molecules that may be present in the compound.

The values of isomer shifts for the D1 and D2 doublets are typical for $Fe^{2+}$ in high spin (S=2) configuration [13]. The minority phase D3 doublet associated to $Fe^{3+}$ state (S = 5/2) exhibits significantly lower isomer shift and quadrupole splitting [13].

*3.3.3 Line widths*

Evidently the width of the lines depends on the presence of water in the studied compound. The doublets associated with the dried sample have the broadest lines. This effect may be related to defects possibly created by removing water. Furthermore, it is clear that the linewidths of the D1 component are the narrowest, while those of the D2 and A3 sub spectra are wider roughly by a factor of 2. Moreover, the widths of D1 and D3 components raise with decrease of water content, what can be related to changes in distribution of Fe ligands. As in case discussed in section 3.3.1, those changes seem to originate from evolution of the triclinic structure during sample drying.

**4. Concluding remarks**

Water content in Fe-gluconate, $Fe(C_6H_{11}O_7)_2 \cdot xH_2O$, has significant effect on its crystallographic structure. Dry sample ($x$=0) crystallizes in the triclinic unit cell (space group *P1*), while fully hydrated sample ($x$=2) exhibits the monoclinic unit cell (space group *I2*). It was established that the partially hydrated sample was a superposition of two phases, *P1* and *I2*, however, the triclinic structure changes during drying process.

The Fe-gluconate (dry and hydrated) shows a complex behavior of lattice parameters versus temperature. However, both samples behave very differently. The hydrated specimen shows smooth thermal expansion of all parameters, apart from the b-parameter. The dry sample exhibits a strong decrease of the b and c lattice parameters with increase of the temperature. The hydrated Fe-gluconate loses water at 100°C, while decomposition of the crystal structure takes place above 150°C.

Although the content of water changes the structure significantly, the Mössbauer spectroscopy confirmed that the $Fe^{2+}$ high spin (S=2) configuration has the highest contribution to the collected data for all specimens. On the other hand, the abundance of the $Fe^{2+}$ components (D1 and D2), changes significantly when structure changes from monoclinic *I2* (hydrated) into the triclinic *P1* (dry). As the quadrupole splitting of those component are little different it suggests that the Fe local environment changes when Fe-gluconate is becoming dry. The $Fe^{3+}$ contribution exhibits much less pronounced changes in the analogue situation. Relative amount of the ferric ions ($Fe^{3+}$) in the fully hydrated sample is by 10% higher than the one in the dry sample.

The above results indicate that there is still room for interesting findings in widely used compounds with well-established biomedical and industrial applications.